\documentclass[]{aastex631}

\newcommand{\kms}{\ensuremath{\rm km\,s^{-1}\ }}
\newcommand{\kmse}{\ensuremath{\rm km\,s^{-1}}}

\received{May 13, 2025}
\revised{}
\accepted{}

\submitjournal{PASP}

\shorttitle{Trumpler's Radial Velocities}
\shortauthors{Trumpler et al.}

\begin{document}

\title{Trumpler's Radial Velocities of  \\ 
Stars in Galactic Star Clusters \footnote{Based on observations at Lick and Keck Observatory.}}

\correspondingauthor{Geoffrey Marcy}
\email{geoff.w.marcy@gmail.com}

%%\author{Robert J. Trumpler $\dagger$} 
%%\email{$\dagger$ \ deceased \ September\ 10, 1956 }

%%\author{Harold Weaver $\dagger\dagger$} 
%%\email{$\dagger\dagger$ \ deceased \ April\ 26, 2017 }
\author{Geoffrey W. Marcy}
\affiliation{Center for Space Laser Awareness, 3388 Petaluma Hill Rd, Santa Rosa, CA 95404}

\author{Dylan M. Lynn}
\affiliation{Department of Astronomy, University of California, Berkeley, CA 94720}
%%\email{dylanlynn.edu@gmail.com}

%%\email{geoff.w.marcy@gmail.com}

\begin{abstract}
We assess the accuracy of radial velocities of 671 stars located near Galactic clusters, measured by Robert Trumpler between 1924 and 1947 using the Lick Observatory 36-inch refractor equipped with prism spectrometers and photographic plates. We find that Trumpler’s velocities share the same zero-point and scale as modern IAU radial velocity standards. Their accuracy ranges from 2 to 7 km/s, depending on stellar brightness, based on comparisons with new Doppler measurements from the Keck telescope and data from SIMBAD. We also provide improved star identifications, B-band photometry, and notes. A link is provided to an online PDF—previously available only in print—containing all 3,782 of Trumpler’s stellar radial velocities and associated Julian dates. These historical data can be combined with modern measurements to form a unique century-long baseline for detecting long-period stellar companions, gravitational perturbations from passing massive objects, and accelerations due to cluster and Galactic dynamics.
 
\end{abstract}


\section{Introduction} \label{sec:intro}

From 1924 to 1947, Robert J. Trumpler obtained spectra and measured the Doppler shifts of 773 stars located near 73 Galactic open clusters, as reported in detail by \citet{Weaver1966,Weaver1967}.  The over-arching motivation was to measure the structure, kinematics, and stellar constituents of the Milky Way Galaxy, which were just being gleaned in the 1920's. The goals of obtaining radial velocities of stars in open clusters included establishing the average velocities and distances of the clusters.  The probability of the true membership of each star located apparently near a cluster could be assessed by how similar its radial velocity was relative to the other members. Additional goals of the spectroscopy were to determine the spectral type and luminosity class of each star and to discover binary stars by their changing Doppler shifts.  The overall motion of each entire open cluster gave vital information about the large-scale differential rotation of the Galaxy.  

However, Trumpler's radial velocity measurements of the stars in open clusters, along with other fundamental properties, were published only in paper form, consisting of 392 pages in three spiral-bound volumes, in \textit{Publications of the Lick Observatory, Volume XXI} \citep{Weaver1966}. Paper copies presumably reside at a few observatories that were active in the 1960's, but we don't know of any besides the one we have. Trumpler's measurements were assembled in this publication by \citet{Weaver1966} due to the untimely death of Robert Trumpler on September 10, 1956.   

Here, we determine the accuracy of these radial velocities compared to the modern IAU radial velocity standard stars. We selected 671 stars, of the 773 in total, that had unambiguous identifications, enabling careful cross-referencing  with modern measurements. We provide an electronic tabulation of the average radial velocities of those 671 stars, along with modern star identifications, B magnitudes, and notes.  We also provide a link to an online PDF of the original 392-page document \citep{Weaver1966}.  Importantly, this PDF contains Trumpler's individual radial velocities and their Julian Dates of observation between 1924 and 1947, providing a time stamp and time baseline for the velocities of those stars.  

\section{The Catalog and Star Identifications} \label{sec:catalog}

A paper copy of the \textit{Publications of the Lick Observatory, Volume XXI} \citep{Weaver1966} was given to us in 2010 by Harold Weaver at U.C. Berkeley.  He requested that we write this paper to assess the quality of Robert Trumpler's radial velocity measurements and to provide an online, searchable portal to their existence and value.  Weaver's paper copy will henceforth reside at the Lick Observatory on Mount Hamilton along with the original photographic plates obtained by Robert Trumpler.  

\begin{figure}[ht]
\plotone{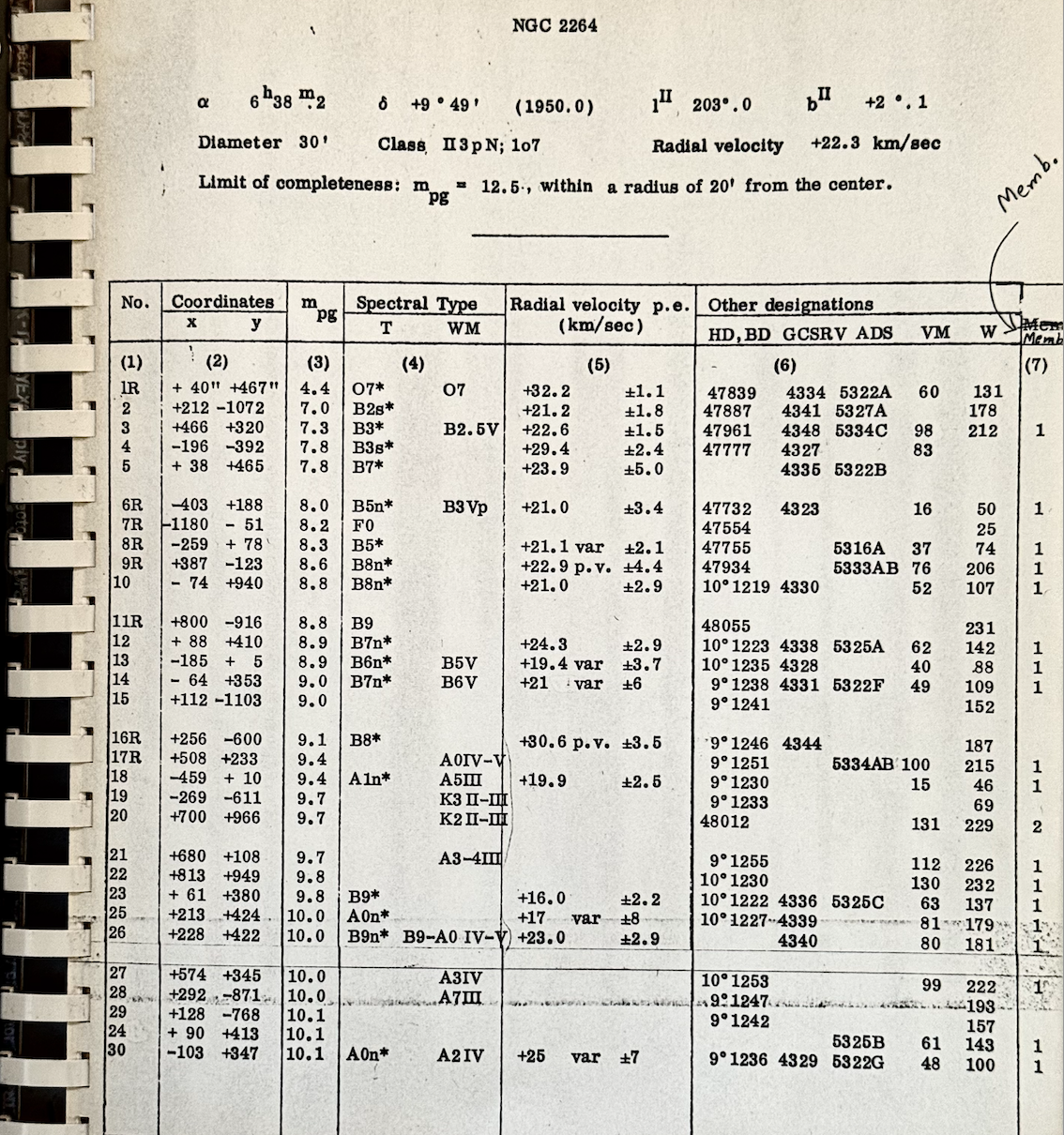}
\caption{A representative page from the three-volume notebook of radial velocity measurements by Trumpler, assembled by \citet{Weaver1966}.  The first column gives Trumpler's internal number for the star within its cluster, marked on a separate photograph. The 2nd and 3rd columns provide the displacement in x and y, in arcseconds, relative to the center of the open cluster, NGC 2264. The next two columns give Trumpler's measured "photographic magnitude" and  spectral type. The next column gives Trumpler's measured radial velocity and probable error, "p.e.". The next column, "Other designations" give various names in catalogs, notably the HD, BD, and General Catalog of Stellar Radial Velocities, and others. The last column is a note about the likely membership in the cluster.}
\label{fig:notebook_page}
\end{figure}

Figure \ref{fig:notebook_page} shows a representative page from that catalog, the first page for the open cluster, NGC2264.   The left column gives Trumpler's personal star number for each star, which is also marked next to its star image on a photograph of the host open cluster, creating a visual identification.  In the next column, Trumpler provided his measured angular distance from the nominal cluster center (noted at the top of the page) in both the x and y axes in arcseconds (RA, DEC). The next two columns give Trumpler's measured "photographic magnitude" (roughly similar to a modern B band) and spectral type.  The next column gives Trumpler's measured heliocentric radial velocity and its "probable error" in \kmse.  

The last column in Figure \ref{fig:notebook_page} from the catalog gives any "Other designations" of the star name, often the HD, BD, and GCSRV \citep{GCSRV}. For stars south of Declination -18, the CPD number from the Cape Photographic Durchmusterung \citep{Gill_Kapteyn1896} is often listed. Roughly 75\% of the stars in the notebook have such an "Other designation", allowing them to be investigated within online repositories of astronomical data, such as SIMBAD.  For stars without an "Other designation" we identified them by using the Trumpler's photographic image of its cluster and his annotated assigned number.  Trumpler provided an image of all 73 open clusters, annotated with Trumpler's assigned numbers for all the stars in the catalog \citep{Weaver1966}. We matched each star in his images to the corresponding star in a sky image from SIMBAD's Aladin Applet and SKY-MAP. 

Further identification of stars was possible with the listed angular separation (column 2), in both RA and DEC in arcseconds, from the coordinates of the nominal center of the open cluster, specified at the top of the page. Trumpler used the forward-looking 1950 equinox for his coordinates.  From those annotations in the photograph and the delta coordinates, we double-checked the identification of each star, including those without an "Other designation". Further, Trumpler's estimated magnitude on the "photographic magnitude scale" is given in column 3 and spectral type in column 4, allowing us to check the identification of all 671 stars in the SIMBAD database. We established a name that appears in SIMBAD for every star.

\section{Trumpler's Radial Velocity Measurement Technique} 

The spectroscopic techniques and Doppler analyses employed by Trumpler were described carefully by \citet{Weaver1966,Weaver1967}. In brief, Trumpler made all observations with the 36-inch refracting telescope at Lick Observatory from 1924 to 1947.  During those years, six different spectrographs were used, each having a different prism or camera focal length, yielding spectral dispersions that ranged from 38 \AA/mm to 132 \AA/mm, measured at 4300 \AA, implying a spectral resolution of 0.1 \AA \ to 0.3 \AA. The wavelengths along the spectrum were calibrated by using a lamp that shined its light through the entrance slit of the spectrograph, with its spectral lines landing above and below the stellar spectrum. Those emission lines, having known laboratory wavelengths, provided the wavelength scale with which to measure the Doppler shift of the spectral lines in the star's spectrum.  During the 23 years, both He lamps and Fe sparks were used to produce the series of comparison emission lines. 

The photographic plates containing the spectra were taken to a "measuring engine" at Lick Observatory to measure the positions of both the stellar spectral lines and the comparison-lamp emission lines.  In practice, a plate was attached to the stage of the measuring engine that could be moved slowly, in the direction of wavelength dispersion, by a screw that a person turned by hand to displace the spectrum along its length. An eyepiece allowed the user to see a cross-hair, composed of a strand of spider web, with the photographic spectrum visible behind it.  While moving the spectrum, the person would notice when a spectral line was centered on the cross-hair and record manually on paper the displacement of the stage calibrated by the screw. In that way every spectral line had a measured position, analogous to a modern pixel location, along the dispersion direction of the spectrum. Trumpler employed assistants to perform these manual measurements of the spectral lines, and he established empirically the systematic biases of each assistant.  

For each spectral type, only certain absorption lines were measured, and Trumpler determined and accounted for the systematic errors caused by blends of multiple stellar absorption lines as a function of star's spectral type and luminosity class. Trumpler corrected the resulting Doppler shifts for many other systematic effects such as from the mechanical flexure of the spectrograph toward different regions of the sky, the temperature and focus of the spectrometer, and the changing curvature of the spectral lines caused by the prism.  Trumpler computed the weighted average of the Doppler shifts from typically 15 to 20 stellar absorption lines to establish a final radial velocity, and he computed their variance to establish an internal uncertainty. 

Trumpler corrected the radial velocities to account for the orbital motion of the Earth around the Sun and for the rotational motion of the telescope about the center of the Earth.  The resulting radial velocities, listed in Table 1, are "heliocentric", i.e. they are the component of the relative velocity vector along the line-of-sight between the star and the Sun.  The Sun changes velocity by $\sim$0.013 kms due to the gravitational pull on the Sun by Jupiter and all the planets, an effect Trumpler properly ignored as less than the errors. Trumpler obtained multiple spectra of some of the stars within the open clusters, offering an opportunity to measure internal errors and to detect single-line and double-line spectroscopic binaries.  Any such variation is noted in Table 1, described below.

\begin{figure}[ht]
\plotone{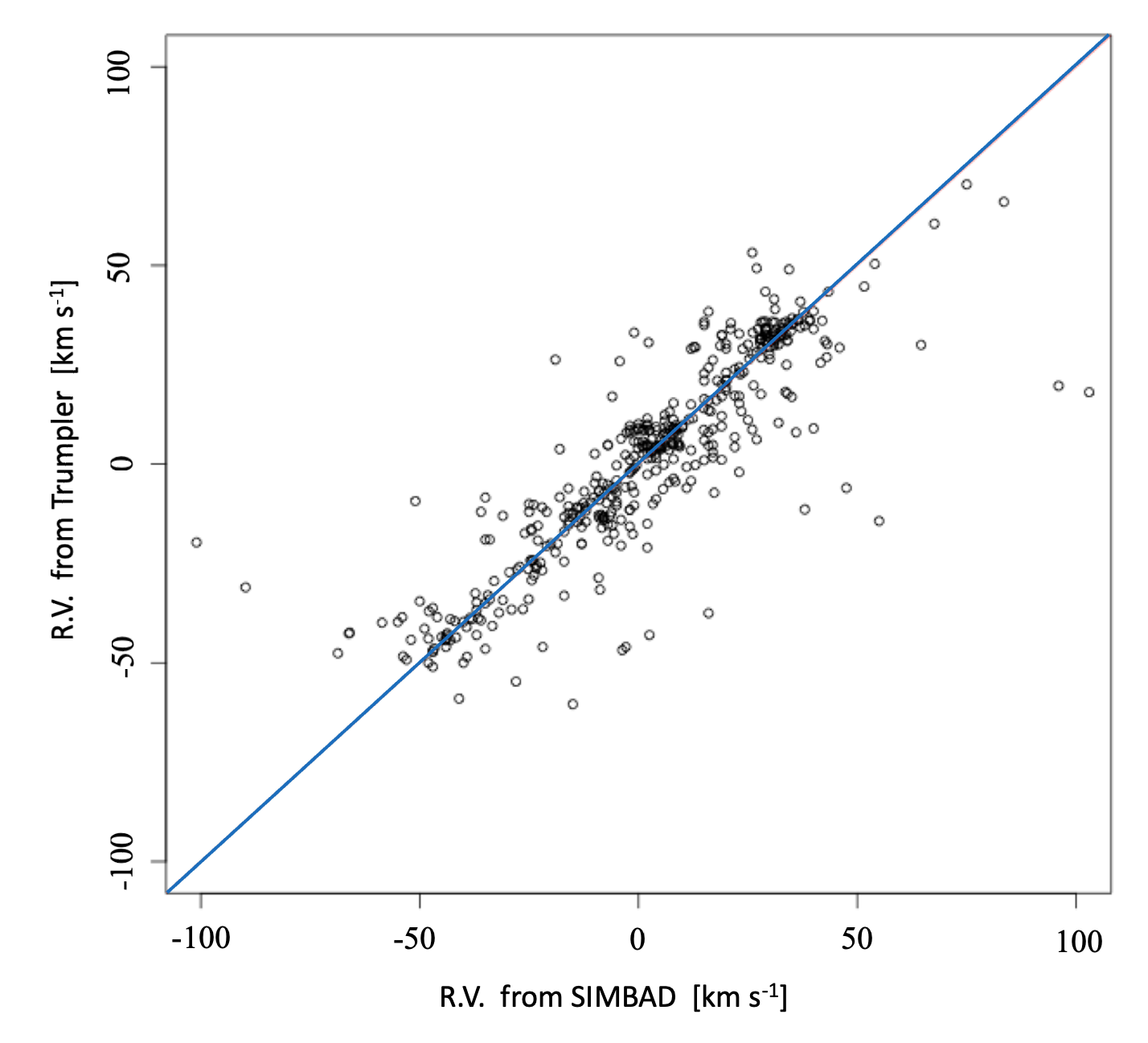}
\caption{Radial velocities from Trumpler vs. radial velocities listed in SIMBAD, for those stars having measurements from both. Trumpler obtained all velocity measurements between 1924 to 1947 while the quoted radial velocities in SIMBAD are more recent. The diagonal blue line represents the locus at which both velocities are the same. The clustering around the diagonal line indicates that Trumpler's radial velocities have the same scale and zero-point as those listed in SIMBAD. The points farthest from the diagonal may be errors in one radial velocity or the other, or they may be due to binary motion during the past $\sim$80 yr.  The standard deviation is 6 \kmse, indicative of the quadrature sum of uncertainaties and any motion in a binary star system.}
\label{fig:general}
\end{figure}

\section{The Measured Radial Velocities}

Table 1 contains 14 columns of data following a format described below. Columns 1 through 5 list multiple standard star names including HD, BD, HIP, GCSRV from the \textit{General Catalogue of Stellar Radial Velocities} \citep{GCSRV}, or "Other ID" such as from Cape Photographic Durchmusterung (CPD) \citep{Gill_Kapteyn1896}.  For the 25\% of the stars, we established a modern star identification, as described above, often using the photographic images of the cluster provided in \citet{Weaver1966} compared to SIMBAD's Aladin Applet and SKY-MAP. Column 6 lists the name of the open cluster that each star was near. Column 7 contains the "photographic magnitude", accurate only to 0.1 mag, for each star measured by Trumpler from the photographic plates of the cluster, obtained by one of the small telescopes at Lick Observatory. Column 8 has the corresponding B magnitude provided by SIMBAD or SKY-MAP as a verification for positive identification of a given star. 

Columns 9 and 10 are the average of the radial velocities measured by Trumpler with his associated calculation of the "probable error" (p.e.). Many stars had multiple radial velocities measurements, and we list here only the average given in \citet{Weaver1966}. Column 11 is the radial velocity provided by SIMBAD when available. Columns 12 and 13 are the Julian Dates (JD-2,400,000) that Trumpler first began taking measurements for a given star and when he last took measurements. The last column, 14, gives the number of the Note, if any, found for a star given by Robert Trumpler, Harold Weaver, SIMBAD's star descriptions, and any other comments discovered while identifying the star names.  The notes themselves are listed after Table 1.

\section{The Uncertainties of the Radial Velocities}

For stars brighter than $\sim$8th mag, Trumpler compared his radial velocities with those in the General Catalogue of Stellar Radial Velocities \citep{GCSRV}. Trumpler found a systematic difference between his raw radial velocities and those in that Catalogue, and the difference depended systematically on spectral type and luminosity class.  He computed the average difference in radial velocity, for a given spectral type and luminosity class, and applied them to all of his raw measurements to yield final radial velocities that had the systematic difference removed between his velocities and those of the Catalogue.  In that way, he ensured that his radial velocities had the same zero-point and scale as those in that General Catalogue.

Trumpler estimated the random errors in his radial velocities by comparing them to those velocities listed in the General Catalogue of Stellar Radial Velocities \citep{GCSRV}.  He found the random errors of his radial velocities to be 2 to 6 \kms (RMS), depending on brightness, spectral type, luminosity class, and the widths of the absorption lines due to rotation, turbulence, and collisional broadening.  

 To test the zero-point and uncertainties of Trumpler's velocities, we took Keck-HIRES spectra on July 1, 2010 of four stars that were measured by Trumpler.  The four stars, all FGK giants and supergiants of roughly 6th magnitude, were HD 172365, HD 160371, HD 161622, and HD 162587.  We measured the velocities from Keck-HIRES spectra using the algorithm written by \citet{Chubak2012} that determines radial velocities in the frame of the solar system barycenter with an accuracy of 0.1 \kms.   The four velocities from Keck-HIRES are -20.21 \kmse, -8.72 \kmse, -6.80 \kmse, -16.16 \kmse, respectively.  Trumpler's velocities for those four stars were -20, -7.4, -7.1, -19.3 \kmse, respectively.   

The mean of the differences between Trumpler's radial velocities and those from Keck-HIRES was -0.48 \kms with a standard deviation of 1.89 \kms. Apparently Trumpler's radial velocity errors for the brightest stars are indeed  $\sim$2 \kmse.   The velocities from \citet{Chubak2012} are consistent with, and helped establish, the zero point and scale of the IAU radial velocity standard stars \citep{Stefanik1999}.  Thus Trumpler's radial velocities seem to have a zero-point consistent with the IAU standards, i.e. no systematic offset, within 1 \kmse, .  

We also compared Trumpler's measured radial velocities to those listed in the SIMBAD database.   We found 450 stars in common.  Figure \ref{fig:general} shows Trumpler's radial velocities plotted against the radial velocities listed in SIMBAD. The diagonal blue line shows the locus of points at which Trumpler's velocities would agree with those in SIMBAD.  The points indeed scatter about the diagonal line, indicating that Trumpler's velocities in Table 1 are on the same velocity scale as those in SIMBAD, with no significant zero-point difference.  The standard deviation is 6 \kmse, some of which is due to binary stars and to the errors and inhomogeneous sources of the SIMBAD velocities. Indeed, the listed radial velocities in SIMBAD change by many \kms over time, no doubt caused by newly injested radial velocities.  \textbf{Thus we conclude that Trumpler's listed radial velocities are consistent with the zero-point and the scale of the IAU radial velocity standard stars, with errors of 2 to 6 \kmse.}

\newpage

\section{Conclusions}

This paper provides the average radial velocities of 671 stars located near Galactic open clusters that were measured by Robert Trumpler from 1924 to 1947.  These radial velocities had existed only in paper form, typed by hand \citep{Weaver1966}, that will now reside at the Lick Observatory.  A further key goal was to provide secure identifications of each of the 671 stars so that the radial velocities, spectral types, and magnitudes could be employed in further research of those stars.  

We performed two checks of the accuracy of Trumpler's radial velocities by comparing them to new radial velocity measurements of four stars observed with the Keck-HIRES spectrometer and to 450 stars having radial velocities listed on SIMBAD. The tests show that Trumpler's radial velocities have a zero-point and a scale consistent with the IAU radial velocity standard stars \citep{Stefanik1999}.  The uncertainties of Trumpler's radial velocities range from 2 \kms for the stars brighter than 8th mag up to 6 \kms for the fainter stars up to 13th mag.  Thus, Trumpler's velocities offer information about membership in their respective Galactic open clusters, and offer a time baseline of nearly a century to search for accelerations caused by binary star companions, cluster dynamics, or exotic rare events such as passing compact objects.  

Finally, we scanned the entire \textit{Publications of the Lick Observatory, Volume XXI} \citep{Weaver1966} and placed a PDF of it online at a location to be specified in publication proofs.  In addition a copy is here:

\url{https://www.dropbox.com/scl/fi/mqk4lc6pviuh9s8u7nfik/PubLickObsVolXXI1966.pdf?rlkey=2uktcyhl1pw5791pp8yedqhp9&st=lfh8y07g&dl=0}

This full copy makes all of its detailed information available online to the astronomical community and to the public. This public version contains the individual radial velocity measurements and associated Julian Dates of each observation, suitable for the detection and orbital analysis of long-period binary stars and for detecting accelerations from any gravitational dynamics.

\section{Acknowledgments}
\begin{acknowledgments}
We thank Harold Weaver for requesting that this paper be written. We also thank him for his tireless work to organize all of Robert Trumpler's radial velocities and other stellar data.  We are deeply appreciative of the careful and voluminous radial velocity measurements by Robert Trumpler, made during three decades, that provide fundamental information about the membership, properties, binary nature, and dynamics of stars in 73 open clusters in our Milky Way Galaxy. Carly Chubak wrote the computer code we used to compute radial velocities in the frame of the solar system barycenter from the Keck-HIRES spectra.  We thank Nicholas Suntzeff, Peter McCullough, Merle Walker, Alex Filippenko, and Ben Reaves for helpful comments about the manuscript. This research made use of the SIMBAD database, operated at CDS, Strasbourg, France.  SIMBAD provided coordinates and photometry enabling identification of roughly half of the stars listed here, along with radial velocity measurements. Some of the data presented herein were obtained at the W. M. Keck Observatory, which is operated as a scientific partnership among the
California Institute of Technology, the University of California, and
NASA; the observatory is made possible by the generous financial support of the W. M. Keck Foundation. 
\end{acknowledgments}

\vspace{5mm}
\facilities{36-inch Refractor Telescope at Lick Observatory, Keck 1 Telescope and HIRES.}

\begin{longrotatetable}
%%
\end{longrotatetable}

\section{Notes from Weaver (1966) and SIMBAD}

Below are the notes found for each star, numerically identified in the last column of Table 1.

{\bf 1 } Star in double system, spectral type and radial velocity taken from GCSRV \citep{GCSRV}.

{\bf 2 } Emission line star.

{\bf 3 } Star in Cluster. Spectroscopic binary orbit. Var.

{\bf 4 } Be star. H-beta, H-gamma emission H-delta not visible. Prob.Var.

{\bf 5 } Emission-line Star. 

{\bf 6 } Star in double system. Var.

{\bf 7 } H-beta - H-delta emission. Companion of magnitude 12, separation 3'', position angle 160 degrees. Data refer to combined light.

{\bf 8 } Emission-line Star. Magnitude refers to light from both components, other data to brighter component. Separation 3''.5, position angle 90 degrees.

{\bf 9 } Star in double system. Var.

{\bf 10 } Star in double system. 

{\bf 11 } Variable star.

{\bf 12 } Variable Star of gamma Dor type .

{\bf 13 } Star in double system.

{\bf 14 } Var.

{\bf 15 } Star in double system.

{\bf 16 } Be star. Double star not noted in ADS, separation approximately 1''. Radial velocity and magnitude refers to light of both components.

{\bf 17 } No standard name, Oja 1966 star No.1164.

{\bf 18 } No standard name. Lines appear double on some spectrograms. Var.

{\bf 19 } Star in double system.

{\bf 20 } Star in double system.

{\bf 21 } Emission-line star.

{\bf 22 } Emission-line star.

{\bf 23 } Be star.

{\bf 24 } Companion of magnitude 12, separation 2''.7, position angle 197 degrees.

{\bf 25 } Prob.Var. 

{\bf 26 } Spectroscopic binary. Var.

{\bf 27 } Spectroscopic binary.

{\bf 28 } Spectroscopic binary. 

{\bf 29 } Spectroscopic binary. 

{\bf 30 } Variable Star. 

{\bf 31 } Var.

{\bf 32 } Var.

{\bf 33 } Data refer to the brighter component, for the combined light magnitude = 10.3.

{\bf 34 } Variable Star of irregular type. 

{\bf 35 } Star in double system. No companion at the position given in the ADS is seen on the photograph.

{\bf 36 } Var.

{\bf 37 } Cepheid variable Star. Radial velocity taken from GCSRV. 

{\bf 38 } Variable Star of beta Cep type. Radial velocity taken from GCSRV.

{\bf 39 } Be star.  

{\bf 40 } Be star.  

{\bf 41 } Variable Star of beta Cep type.  

{\bf 42 } Be star.  

{\bf 43 } Var.

{\bf 44 } Be star. H-beta,  H-gamma  in emission; all absorption lines faint.

{\bf 45 } Variable Star of beta Cep type. Some lines appear double on one spectrogram. 

{\bf 46 } Spectral type and radial velocity from GCSRV.

{\bf 47 } Semi-regular pulsating star.

{\bf 48 } Variable star.

{\bf 49 } Variable star of beta Cep type.

{\bf 50 } Double line spectroscopic binary orbit. Var.

{\bf 51 } Eclipsing binary Algol type. Var.

{\bf 52 } Var.

{\bf 53 } H-beta faint emission; H-gamma weak absorption. Be star. Var.

{\bf 54 } H-beta and H-gamma emission on broad absorption, Be star.

{\bf 55 } H-beta emission; H-gamma, H-delta not visible; poor lines. Be star.

{\bf 56 } Variable star of beta Cep type.

{\bf 57 } Spectral type and radial velocity from GCSRV, revised in accordance with additional observations communicated by R.M. Petrie.  Semi-regular pulsating star.

{\bf 58 } H-beta H-epsilon emission

{\bf 59 } Spectral type and radial velocity from GCSRV.  Pulsating variable star.

{\bf 60 } Semi-regular pulsating star.

{\bf 61 } Spectroscopic binary orbit. Var.

{\bf 62 } Strong broad emission at 4684 Ang. and  faint emission at 4640 Ang., Emission-line star. Var.

{\bf 63 } Spectroscopic binary orbit. Var.

{\bf 64 } Spectroscopic binary orbit. Var.

{\bf 65 } Double lines noted on one spectrogram. Prob.Var.

{\bf 66 } Prob.Var.

{\bf 67 } Star in association.

{\bf 68 } Star in double system, ADS 1920B

{\bf 69 } Prob.Var.

{\bf 70 } Star in double system.

{\bf 71 } Star in double system.

{\bf 72 } Star in double system.

{\bf 73 } Var.

{\bf 74 } Same spectral type.

{\bf 75 } Var.

{\bf 76 } Separation 1''.3, data refer to combined light. Double or multiple star.

{\bf 77 } Star in double system.

{\bf 78 } Star in double system.

{\bf 79 } Star in double system.

{\bf 80 } Star in double system.

{\bf 81 } Orbit determination of triple system. Var.

{\bf 82 } Visual double, equal magnitudes; separation 0''.3, data refers to combined light. Variable star.

{\bf 83 } Trumpler: ADS 2161B,  SIMBAD:ADS 2161 Double or multiple star.

{\bf 84 } Var.

{\bf 85 } Star in double system.

{\bf 86 } Star in double system.

{\bf 87 } Star in double system. Prob.Var.

{\bf 88 } Be star. name Alcyone. Trace of H-beta emission on some spectrograms.

{\bf 89 } Spectroscopic binary, name Atlas. Doubtful if close double. 

{\bf 90 } Be star, name Electra.

{\bf 91 } Variable star, name Maia.

{\bf 92 } Be star, "Merope".

{\bf 93 } Variable star, name Taygeta.

{\bf 94 } Be star, name Pleione. Variable spectral class; photoelectric variable. 

{\bf 95 } Variable star, HR 1172.

{\bf 96 } Variable star, name Celeno.

{\bf 97 } 18 Tau.

{\bf 98 } Variable star, name Asterope.

{\bf 99 } Star in double system, 24 Tau.

{\bf 100 } 22 Tau.

{\bf 101 } Star in double system.  

{\bf 102 } Variable star with rapid variations. Data refer to brighter component.

{\bf 103 } Spectroscopic binary.  Data refer to brightest component only. Var.

{\bf 104 } Eclipsing binary of Algol type, V* V1229 Tau.

{\bf 105 } Prob.Var.

{\bf 106 } Spectroscopic binary.   

{\bf 107 } Physical membership doubtful on the basis of proper motion. But from magnitude-spectral type relation and radial velocity it appears to be a member.

{\bf 108 } Star in double system. Our 6 radial velocity observations alone would indicate variable velocity; if GCSRV observations are included, however, variability becomes doubtful.

{\bf 109 } Variable star of delta Sct type, V* V1228 Tau.

{\bf 110 } Variable star of delta Sct type, V* V650 Tau.

{\bf 111 } Close double, data refer to combined light.

{\bf 112 } V*V1187 Tau, Variable star of delta Sct type.

{\bf 113 } V*V624 Tau, Variable star of delta Sct type.

{\bf 114 } V*V647 Tau, Variable star of delta Sct type.

{\bf 115 } V*V534 Tau, Variable star of delta Sct type. Close double, data refer to combined light.

{\bf 116 } Variable star of delta Sct type.

{\bf 117 } Var.

{\bf 118 } HR 1185.

{\bf 119 } Variable star. 

{\bf 120 } HR 1103, Variable star.

{\bf 121 } Spectroscopic binary. Radial velocity seems to indicate this star is not a member.

{\bf 122 } Star in double system. Wide double, Boss GC 1852;  data refer to brighter component.

{\bf 123 } Data refer to brighter component.

{\bf 124 } V*V1225 Tau, Variable star of gamma Dor type.

{\bf 125 } Membership somewhat doubtful; neither proper motion nor radial velocity agree very closely with the motion of the cluster.

{\bf 126 } \citet{GCSRV} apparently does not refer to this star.

{\bf 127 } Variable star.

{\bf 128 } Star in double system.

{\bf 129 } Spectroscopic binary, radial velocity from GCSRV. Var.

{\bf 130 } V*SZ Cam, Eclipsing binary of beta Lyr type. Eclipsing and spectroscopic binary, radial velocity from GCSRV. Var.

{\bf 131 } Star in double system. Prob.Var.

{\bf 132 } Variable star. Designations confused in ADS.

{\bf 133 } Variable star. Designations confused in ADS.

{\bf 134 } Star in double system.

{\bf 135 } Designations confused in ADS.

{\bf 136 } Star in double system. Designations confused in ADS. Prob.Var.

{\bf 137 } Prob.Var.

{\bf 138 } Spectroscopic binary orbit. Var.

{\bf 139 } ADS 3074, separation 0''.3; data refer to combined light.

{\bf 140 } No standard name, spectral type B9.

{\bf 141 } HR 1517 Spectroscopic binary. Radial velocity from GCSRV.

{\bf 142 } Emission line star. H lines broad and diffuse; H-beta, H-gamma show faint emission at center. Var.

{\bf 143 } Star in double system

{\bf 144 } Prob.Var.

{\bf 145 } Var.

{\bf 146 } Prob.Var.

{\bf 147 } Spectral type not in SIMBAD (Trumpler: B4*).

{\bf 148 } Spectral type not in SIMBAD (Trumpler: B7*). Var.

{\bf 149 } Prob.Var.

{\bf 150 } Star in double system, Spectral type not in SIMBAD (Trumpler: B9*).

{\bf 151 } Spectral type not in simbad (Trumpler: A0*).

{\bf 152 } ADS 4192A; separation 4''.3; brighter component. Star in double system.

{\bf 153 } V*V1046 Ori, Spectroscopic binary. Var.

{\bf 154 } Variable star. 

{\bf 155 } ADS 4176AB; separation 1''.9; components nearly equal in magnitude. Data refer to combined light. Var.

{\bf 156 } ADS 4172A; separation 1''.2. Radial velocity and spectral type (T)  refer to brighter component, magnitude to combined light.

{\bf 157 } Variable star. Prob.Var.

{\bf 158 } Some lines appear double on one spectrogram. Prob. Var.

{\bf 159 } ADS 4192B.

{\bf 160 } Variable star. Prob.Var.

{\bf 161 } Spectroscopic binary orbit. Var.

{\bf 162 } Prob.Var.

{\bf 163 } Star in double system. Prob.Var.

{\bf 164 } Star in double system. Spectroscopic binary orbit. Var.

{\bf 165 } Some lines appear double on one spectrogram.

{\bf 166 } Emission line star. H-beta ñ H-epsilon emission on broad absorption.

{\bf 167 } Star in double system. Var.

{\bf 168 } Has faint companion (magnitude 14.6); separation 5''.8, position angle 82 degrees.

{\bf 169 } Emission line star. Var.

{\bf 170 } Var.

{\bf 171 } Double lines on some plates.

{\bf 172 } Be Star. Prob.Var.

{\bf 173 } Var.

{\bf 174 } Double or multiple star.

{\bf 175 } Variable star of beta Cep type. Var.

{\bf 176 } Star in double system.

{\bf 177 } Perhaps a double line spectroscopic binary. Variable star of beta Cep type.

{\bf 178 } Star in double system.

{\bf 179 } Star in double system.

{\bf 180 } Star in double system.

{\bf 181 } Variable star of alpha2 Cvn type.

{\bf 182 } Variable star RR Lyr.

{\bf 183 } Spectroscopic binary orbit.

{\bf 184 } Star in double system.

{\bf 185 } Possibly: HD252353 BD+24 1125

{\bf 186 } Prob.Var.

{\bf 187 } Prob.Var.

{\bf 188 } NGC 2168 307.

{\bf 189 } Variable star.

{\bf 190 } Emission-line star.

{\bf 191 } Emission-line star.

{\bf 192 } Emission-line star. Var.

{\bf 193 } Eclipsing binary. spectroscopic binary orbit Listed twice in GCSRV as 4183 and 4186. Var.

{\bf 194 } Radial velocity from GCSRV.

{\bf 195 } Star in double system. Var.

{\bf 196 } Star in double system.

{\bf 197 } Star in double system.

{\bf 198 } Star in double system.

{\bf 199 } Prob.Var.

{\bf 200 } Prob.Var.

{\bf 201 } With a change in position angle of 180 degrees' the measurement fits the pair 7-16 quite well. The position given in ADS is off by 5', but there is no double star at the ADS position. Star 16 was also on the  slit during the exposure.

{\bf 202 } Pre-main sequence Star. S Monocerotis.

{\bf 203 } Pre-main sequence Star

{\bf 204 } Pre-main sequence Star

{\bf 205 } Pre-main sequence Star

{\bf 206 } Star in double system

{\bf 207 } Possible ellipsoidal variable star. Some lines appear double on two spectrograms.

{\bf 208 } Eclipsing binary of Algol type. Double line spectroscopic binary, involved in bright nebulosity. Var.

{\bf 209 } Pre-main sequence star. Close visual binary (0''.8), not separated. Prob.Var.

{\bf 210 } Variable star

{\bf 211 } Pre-main sequence star. Trumpler's BD name conflicts with GCSRV. The GCSRV seems to be the star (BD+10 1235). Var.

{\bf 212 } Young stellar object. Var.

{\bf 213 } Some lines appear double on one spectrogram. Prob.Var.

{\bf 214 } Young stellar object. Var.

{\bf 215 } Young stellar object.

{\bf 216 } Young stellar object, Trumpler's BD ADS, and GCSRV were wrong (BD+09 1236 ADS 5322G GCSRV4329). Var.

{\bf 217 } Star in double system.

{\bf 218 } Has close companion; separation 1''.8, position angle 241 degrees. Data refer to combined light. This stat is also listed as ADS 5482A but with an error in position. 

{\bf 219 } Star in double system

{\bf 220 } Prob.Var.

{\bf 221 } Star in double system

{\bf 222 } Double star not noted in ADS; separation 2'', position angle 27 degrees. Star 8, which is 1 magnitude fainter than star 5, was also on the slit during exposures for star 5.

{\bf 223 } Star in double system

{\bf 224 } Spectroscopic binary orbit. Var.

{\bf 225 } Star in double system.

{\bf 226 } Star in double system.

{\bf 227 } Prob.Var. 

{\bf 228 } Eclipsing binary of beta Lyr type. Has companion of magnitude 14 at position angle 99 degrees, separation 4''.6. Var.

{\bf 229 } Young stellar object.

{\bf 230 } Young stellar object.

{\bf 231 } Young stellar object.

{\bf 232 } Young stellar object.

{\bf 233 } Young stellar object.

{\bf 234 } Star in double system. Stars 1 and 2 were both on the slit during the exposures.

{\bf 235 } Star in double system.  

{\bf 236 } Lines appear double on two spectrograms, probably double line spectroscopic binary.

{\bf 237 } Emission-line star.

{\bf 238 } Star in double system.

{\bf 239 } Emission-line star.

{\bf 240 } Trumpler's BD was a different star

{\bf 241 } Trumpler's BD was a different star

{\bf 242 } Be star. H-beta, H-gamma emission.  Prob.Var.

{\bf 243 } Be star.

{\bf 244 } Star in double system.

{\bf 245 } Star in double system.

{\bf 246 } Star in double system.

{\bf 247 } Star in double system. Lines appear double on one spectrogram. Prob.Var.

{\bf 248 } Prob.Var.

{\bf 249 } Star in double system.

{\bf 250 } Star in double system.

{\bf 251 } Eclipsing binary of Algol type. Var.

{\bf 252 } Var.

{\bf 253 } Cl* NGC 2422 PMS 1156 has a B of 9.72 and it's coordinates are closer to the star than the BD Trumpler lists.

{\bf 254 } Var.

{\bf 255 } Star in double system. Double star measurements indicate that this star has the same proper motion as star 1 and is therefore probably a cluster member; the relatively large deviation of the radial velocity may be due to velocity variation, since it is less than 2.5 times the probable error. Var.

{\bf 256 } Semi-regular pulsating star.

{\bf 257 } Star in double system.

{\bf 258 } Emission-line star. H-beta emission on broad absorption; H-gamma faint.

{\bf 259 } Perhaps double line binary. Var.

{\bf 260 } Star in double system. Var.

{\bf 261 } USNOA2 0525-05946134

{\bf 262 } Var.

{\bf 263 } Double star, brighter component.

{\bf 264 } Var.

{\bf 265 } Has companion of magnitude 10 at position angle 81 deg,separation 3''. Magnitude refers to combined-light radial velocity and spectral type to brighter component.

{\bf 266 } Spectral type peculiar: metallic lines strong while Ca\textit{II} H\&K lines resemble those of an A2 star; may be composite.

{\bf 267 } Some lines appear double; perhaps a double line spectroscopic binary.

{\bf 268 } Var.

{\bf 269 } Spectroscopic binary. Var.

{\bf 270 } Star in double system.

{\bf 271 } Variable star of delta Sct type.

{\bf 272 } Spectroscopic binary.

{\bf 273 } Spectroscopic binary.

{\bf 274 } Variable star of delta Sct type.

{\bf 275 } Star in double system.

{\bf 276 } Variable star.

{\bf 277 } Prob.Var.

{\bf 278 } Variable star.

{\bf 279 } Star in double system.

{\bf 280 } Spectroscopic binary. Radial velocity from GCSRV, spectroscopic binary orbit by R.F Sanford. Var.

{\bf 281 } Spectroscopic binary.

{\bf 282 } Variable star.

{\bf 283 } Spectroscopic binary. Var.

{\bf 284 } Variable star.

{\bf 285 } Variable star of delta Sct type.

{\bf 286 } Spectroscopic binary. Var.

{\bf 287 } Variable star of delta Sct type.  \citet{GCSRV} lists this as a double line spectroscopic binary; Trumpler's spectra confirm variability.

{\bf 288 } Variable star of delta Sct type.

{\bf 289 } Variable star.

{\bf 290 } Variable star.

{\bf 291 } Variable star of delta Sct type.

{\bf 292 } Variable star of delta Sct type.

{\bf 293 } Variable star. Var.

{\bf 294 } Variable star of delta Sct type.

{\bf 295 } Variable star. \citet{GCSRV} lists this as a double line spectroscopic binary; our observations confirm velocity variability, but do not show double lines. Var.

{\bf 296 } Variable star.

{\bf 297 } Variable star of delta Sct type.

{\bf 298 } Variable star.

{\bf 299 } Star in double system.

{\bf 300 } Prob.Var.

{\bf 301 } Northern component of double star.

{\bf 302 } Star in double system.

{\bf 303 } Star in double system.

{\bf 304 } Star in double system.

{\bf 305 } Star in double system.

{\bf 306 } Close double star, according to van Maanen; data refer to combined light.

{\bf 307 } Red giant star.

{\bf 308 } Red giant star.

{\bf 309 } Variable star of delta Sct type.

{\bf 310 } Red Giant branch star.

{\bf 311 } Spectroscopic binary. Has two faint companions at 21'', position angle 282 degrees and 18'', position angle 128 degrees. Prob.Var.

{\bf 312 } Eclipsing binary of beta Lyr type. Brownlee and Cox find some variability indicated; double line binary orbit determined by O. Struve. Two of Trumpler's plates show normal (single) lines, two show very broad lines not clearly separated. Var.

{\bf 313 } Spectroscopic binary. Data refer to brighter component.

{\bf 314 } Wolf-Rayet star. Spectroscopic binary orbit determined by O. Struve. The spectrum shows an emission band at 4686 A and several other broad emissions. Var.

{\bf 315 } Var.

{\bf 316 } Eclipsing binary of Algol type.

{\bf 317 } Radial velocity from \citet{GCSRV}.

{\bf 318 } Spectroscopic binary orbit. Var.

{\bf 319 } Eclipsing binary of Algol type. Double line spectroscopic binary orbit. Var.

{\bf 320 } Double star given in SDS; data refer to brighter (s.p.) component. Var.

{\bf 321 } Var.

{\bf 322 } Star in double system.

{\bf 323 } Prob.Var.

{\bf 324 } Be star. H-beta faint emission. Var.

{\bf 325 } Prob.Var.

{\bf 326 } Semi-regular pulsating star. Irregular variable; its physical membership is doubtful.

{\bf 327 } Var.

{\bf 328 } Visual double star, SDS I 608, companion of magnitude 12.3, separation 2''.0, position angle 304 degrees. The magnitude refers to combined light, radial velocity and spectral type to the brighter component. 

{\bf 329 } Spectroscopic binary.

{\bf 330 } Spectroscopic binary. Radial velocity from GCSRV.

{\bf 331 } Spectroscopic binary. Var.

{\bf 332 } Spectroscopic binary.

{\bf 334 } Star in double system.

{\bf 335 } Var. Close double (separation 0''.7). Data refer to combined light.

{\bf 336 } Be star.

{\bf 337 } Prob.Var.

{\bf 338 } Star in double system. Radial velocity observations were made for each component separately; magnitudes refer to combined light. According to ADS 67a is 0.3 magnitude brighter than 67b, which would make the photographic magnitudes of the two stars 9.6 and 9.9. Prob.Var.

{\bf 339 } Star in double system.

{\bf 340 } Prob.Var.

{\bf 341 } Spectroscopic binary. Var.

{\bf 342 } Prob.Var.

{\bf 343 } Star in double system.

{\bf 344 } Rotationally variable star. Prob.Var.

{\bf 345 } Spectroscopic binary.

{\bf 346 } Close double star of 0''.4 separation, observed as one star.

{\bf 347 } Variable star of alpha2 Cvn type.

{\bf 348 } Spectroscopic binary. Var.

{\bf 349 } Close double star 0''.45 separation, observed as one star. The exceptional spectral type and rather large deviation in radial velocity make it doubtful that this star is a physical member of the cluster.

{\bf 350 } Spectroscopic binary.

{\bf 351 } Variable star of alpha2 Cvn type.

{\bf 352 } Spectroscopic binary.

{\bf 353 } Variable star of alpha2 Cvn type. Var.

{\bf 354 } Spectroscopic binary.

{\bf 355 } Prob.Var.

{\bf 356 } Spectroscopic binary.

{\bf 357 } Sky-map B is 9.998.

{\bf 358 } Prob.Var. Double star, n.f. Component.

{\bf 359 } Double star, n.f. Component.

{\bf 360 } Not found in SIMBAD, B from sky-map.

{\bf 361 } Star in double system.

{\bf 362 } Var.

{\bf 363 } Boss GC 24590.

{\bf 364 } Var.

{\bf 365 } Not found in SIMBAD.

{\bf 366 } Emission-line star. Lies outside the limiting circle adopted for the cluster and is probably  not a cluster member. In nearly all O-type stars, the two or three brightest stars are in the central part of the cluster. Radial velocity from \citet{GCSRV}.

{\bf 367 } Spectroscopic binary. Lies outside the limiting circle adopted for the cluster and is probably  not a cluster member. In nearly all O-type stars, the two or three brightest stars are in the central part of the cluster.  Radial velocity from \citet{GCSRV}.

{\bf 368 } Be star. Lines appear double; probably double line binary. H-beta, H- gamma emission. Var.

{\bf 369 } Double line spectroscopic binary orbit. Var.

{\bf 370 } H-beta, H-gamma faint emission on broad absorption; perhaps double lines on some plates. Var.

{\bf 371 } H-beta, H-gamma faint emission on broad absorption. 

{\bf 372 } Line width variable; perhaps double lines. Var.

{\bf 373 } Visual double star, ADS 11024, SDS h5010; for combined light, magnitude  8.9. The large deviation of the radial velocity makes it probable that this star is not a cluster member.

{\bf 374 } H-beta, H- gamma faint emission on broad absorption; perhaps double lines on some plates. Prob.Var.

{\bf 375 } H-beta, H- gamma faint emission on broad absorption. 

{\bf 376 } Eclipsing binary of beta Lyr type. Var.

{\bf 377 } Emission-line star. Prob.Var.

{\bf 378 } Var.

{\bf 379 } Emission-line star.

{\bf 380 } Si IV lines faint. Prob.Var.

{\bf 381 } Si IV lines faint.

{\bf 382 } Var.

{\bf 383 } Some lines appear double in one spectrogram. Prob.Var.

{\bf 384 } Be star. Var.

{\bf 385 } H-beta, H-gamma, H-delta emission on broad absorption.

{\bf 386 } Star in nebula. Radial velocity from \citet{GCSRV}.

{\bf 387 } Sky-map component B is mag 10.2. 

{\bf 388 } Var.

{\bf 389 } Variable star. Radial velocity  from GCSRV.

{\bf 390 } Classical Cepheid (delta Cep type). U Sgr. Spectral type variable from gF8-gG2 Prob.Var.

{\bf 391 } Star in double system. Var.

{\bf 392 } Be star.

{\bf 393 } Radial velocity from GCSRV.

{\bf 394 } Double lines observed on one spectrogram. Prob.Var.

{\bf 395 } Prob.Var.

{\bf 396 } Prob.Var.

{\bf 397 } Emission-line star. H-beta faint emission; perhaps double lines on some plates. Prob.Var.

{\bf 398 } Star in double system. H lines narrow. Northern component of double star. Var.

{\bf 399 } Star in double system.

{\bf 400 } Emission-line star. Close double separation 0''.7; data refer to combined light. H-beta faint emission on some plates.

{\bf 401 } Star in double system.

{\bf 402 } Classical Cepheid (delta Cep type).

{\bf 403 } Outside the limits of our chart. ADS 11526A, has companion of magnitude 10 at separation 1''.2. The radial velocity leaves no doubt that this star is not a cluster member.

{\bf 404 } Eclipsing binary. Var.

{\bf 405 } Prob.Var.

{\bf 406 } Prob.Var.

{\bf 407 } Var.

{\bf 408 } Spectroscopic binary.

{\bf 409 } Si II lines strong.

{\bf 410 } Star in double system. Listed as double line spectroscopic binary, spectral type cA3. Double lines on some plates. Prob.Var.

{\bf 411 } B magnitude from sky-map.

{\bf 412 } Probably not cluster member, radial velocity differs considerably, and is also outside the limiting circle.

{\bf 413 } Prob.Var.

{\bf 414 } Var.

{\bf 415 } Probably not a cluster member; Radial velocity differs considerably and is also outside the limiting circle.

{\bf 416 } Prob.Var.

{\bf 417 } Prob.Var.

{\bf 418 } Probably not a cluster member; the radial velocity differs considerably.

{\bf 419 } Star in double system.

{\bf 420 } Var.

\newpage
\bibliography{Trumpler_RVs}
\bibliographystyle{aasjournal}
%%\bibliographystyle{plain}


\end{document}